\documentstyle[editedvolume,epsf,psfig,graphicx,numreferences]{crckapb} 

\begin{opening}
\title{Neutrino propagation in Neutron Matter 
and the Nuclear Equation of State} 

\author{J. Margueron}
\institute{Groupe de Physique Th\'eorique, IPN, F-91406 Orsay, France}
\author{J. Navarro}
\institute{IFIC, Apdo. 22085, 46075 Valencia, Spain}
\author{N. Van Giai}
\institute{Groupe de Physique Th\'eorique, IPN, F-91406 Orsay, France}
\author{W. Jiang}
\institute{Institute of Nuclear Research, Shanghai 201800, China}
\end{opening}

\runningtitle{NEUTRINO PROPAGATION IN NEUTRON MATTER}

\begin{document}

\noindent
{\bf Abstract} 

\noindent
We study the propagation of neutrinos inside dense matter under the
conditions prevailing 
in a proto-neutron star. Equations of state obtained with
different nuclear effective interactions (Skyrme type and Gogny type) are
first discussed. It is found that for many interactions, spin and/or isospin
instabilities occur at densities larger than the saturation density of
nuclear matter.  
From this study we select two representative interactions, 
SLy230b and D1P.  
We calculate the response functions in pure neutron matter where 
nuclear correlations are described at the Hartree-Fock 
plus RPA level. These response functions allow us to evaluate 
neutrino mean free paths corresponding to neutral current processes. 

\section{Introduction}

Neutrinos play a crucial role during the energy liberation 
phase that follows a super-nova collapse as they carry away
most of the initial gravitational energy. In order to describe
the evolution of the newly born neutron star it is important to
have reliable estimates of the mean free path of neutrinos in a
medium with the typical characteristics of a proto-neutron star,
namely with densities ranging between one and several times the
value of the saturation density of symmetric nuclear matter,
with temperatures of around a few tens of MeV, and 
with a proton fraction which can represent up to 
30\% of the baryonic density. 

We report here on some calculations of the mean free path of
neutrinos in pure neutron matter 
under various conditions of
density and temperature. The scattering of neutrinos on neutrons
is mediated by the neutral current of the electroweak
interaction. In the non-relativistic limit and in the case of
non-degenerate neutrinos, the mean free path of a neutrino with
initial momentum ${\bf k}_1$ is given by~\cite{IP:82}
\begin{eqnarray}
1/\lambda({\rm k}_1,T) &=& \frac{G_F^2}{16 \pi^2} \int d{\bf k}_3  
\Bigg( c_V^2 (1+\cos{\theta})~{\cal S}^{(0)}(q,T) \nonumber\\
&& \hspace{3cm} + c_A^2 (3-\cos{\theta})~{\cal S}^{(1)}(q,T) \Bigg)~,
\label{eq1}
\end{eqnarray}
where $T$ is the temperature, $G_F$ is the Fermi constant, $c_V$
($c_A$) the vector (axial) coupling constant, ${\bf k}_3$ the
final neutrino momenta, $q = k_1-k_3$ the transferred
energy-momentum, and $\cos\theta=\hat{\bf k}_1\cdot\hat{\bf
k}_3$.  The dynamical structure factors ${\cal S}^{(S)}(q,T)$
describe the response of neutron matter to excitations induced
by neutrinos, and they contain the relevant information on the
medium. The vector (axial) part of the neutral current gives
rise to density (spin-density) fluctuations, corresponding to
the $S=0$ ($S=1$) spin channel.

A microscopic framework based on effective nucleon-nucleon
interactions is used here to describe consistently both the
Equation of State (EOS) and the dynamical structure factors of
neutron matter. Then, the mean free path of neutrinos is
calculated according to Eq.~(\ref{eq1}) for different values of
the density and temperature of the nuclear medium. 

\section{The EOS and the choice of effective interactions}
\begin{figure}
\centering
\includegraphics[scale=0.36,angle=-90]{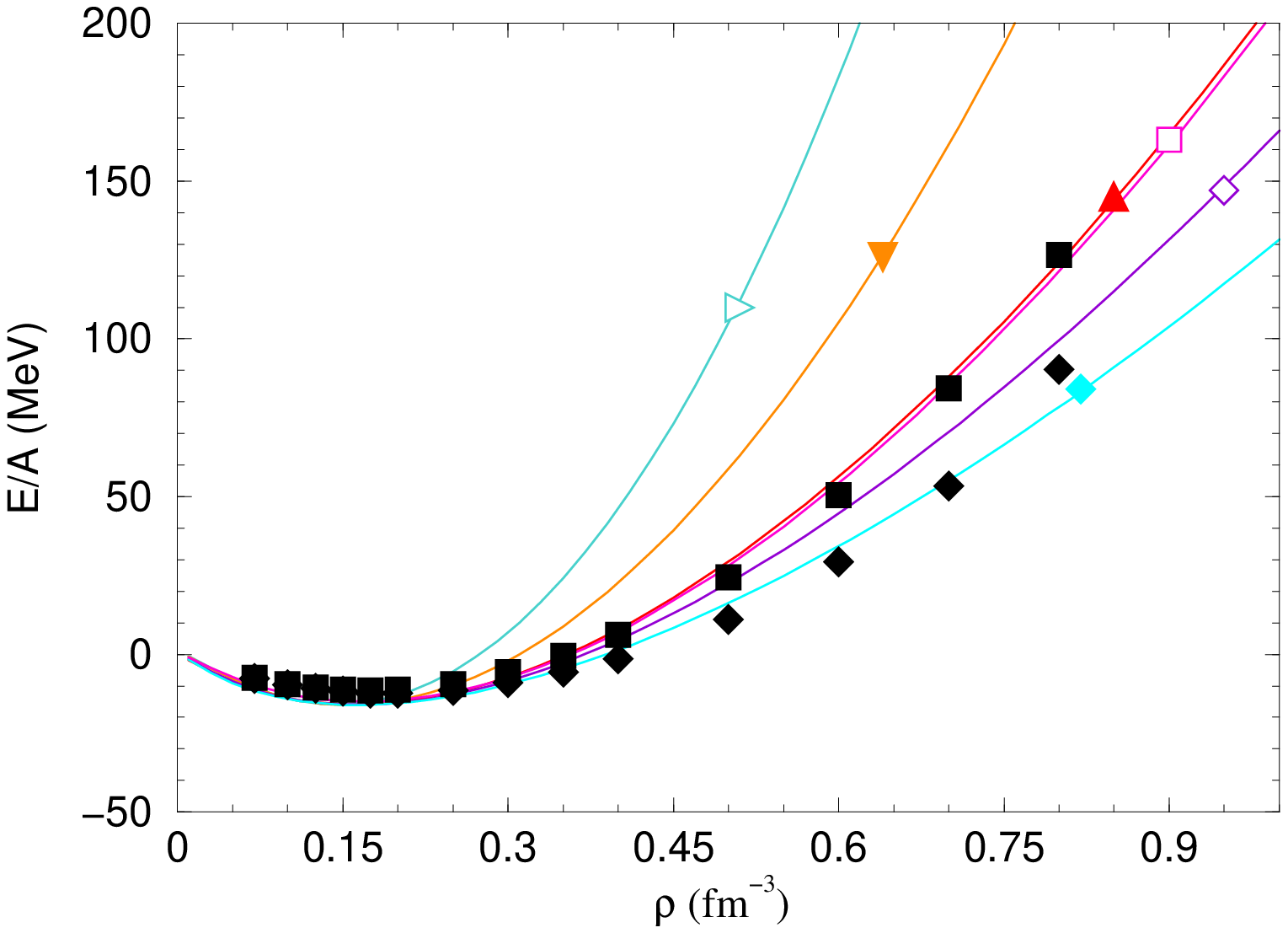}
\includegraphics[scale=0.36,angle=-90]{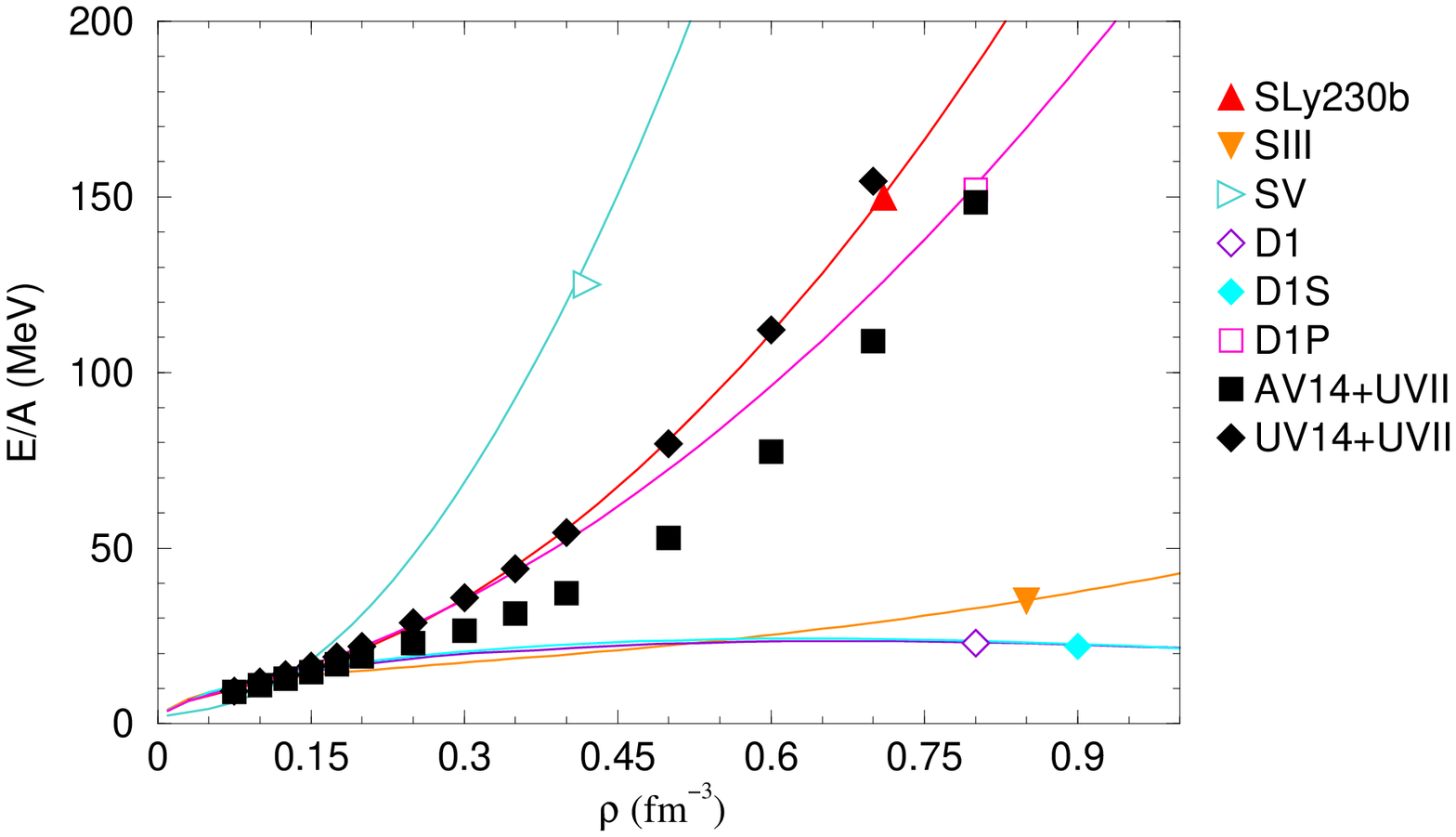}
\caption{Energy per particle of symmetric nuclear matter (left
panel) and neutron matter (right panel) as a function of
density $\rho$ for a selected set of Skyrme and Gogny effective
interactions.  The black squares and diamonds are the results of variational
calculations[2] based on realistic two- and three-body
interactions.}
\label{fig1}
\end{figure}

Nucleon-nucleon effective interactions are usually fitted so as
to reproduce some selected properties of finite nuclei and
symmetric nuclear matter at saturation. The study of
proto-neutron stars requires the use of these interactions for
nuclear densities well beyond the saturation density of symmetric
nuclear matter, and we explain in this section the
criteria for our choice among the plethora of effective
interactions currently in use.  In Figure~\ref{fig1} is
represented the energy per particle in symmetric nuclear matter
and pure neutron matter 
as a function of the density $\rho$, as resulting from a mean field
calculation employing several Skyrme and Gogny interactions.
For comparison, the variational results given in \cite{WFF:88}
are also plotted; such results are based on the two realistic
two-body interactions AV14 and UV14, plus the three-body
interaction UVII. The noticeable differences between these two
upper bounds provide an estimate of the reliability of the
calculations for densities beyond the saturation point.  

In the left panel of Fig.~\ref{fig1} one can see that the
Skyrme forces SIII and SV overestimate $E/A$ 
as compared with the variational results for values of
the density larger than $0.25$ fm$^{-3}$. Such a discrepancy has
to be related to the high value of the compression 
modulus provided by these interactions ($\sim 350$ MeV). The
remaining effective interactions used in this figure give more
reasonable values for the incompressibility (around 240 MeV), and
one can see that their predicted energies per particle are in
reasonable agreement with the variational results in the
range of densities displayed in the figure.  In
contrast, in neutron matter (right panel) only the
effective interactions SLy230b and D1P lead to energies per particle in
agreement with the variational results. As a matter of fact, both SLy230 and
D1P have 
included the variational results of neutron matter 
into the adjustment of parameters. 
Thus, we select two representative forces, one of Skyrme type
(SLy230b\cite{CHA97}) and one of Gogny type (D1P\cite{FVSBDG:99}).  

\begin{figure}
\centering
\includegraphics[scale=0.4,angle=-90]{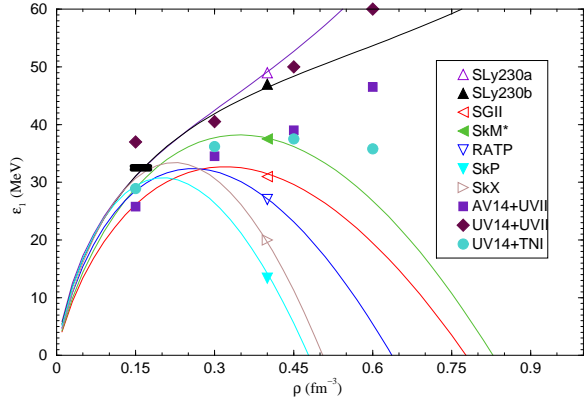}
\caption{Asymmetry energy $\epsilon_1$ of nuclear matter 
as a function of density for a
set of Skyrme and Gogny effective interactions. We also show the
results of variational calculations[2]. The black rectangle
represents 
the empirical point including error bars.}
\label{fig3}
\end{figure}

However, the agreement with the variational EOS is not the only
criterium we have followed to select an effective interaction.
We have also considered isospin and spin properties. The
asymmetry energy is related to the second derivative of the
energy per particle with respect to the asymmetry parameter  
$I=(\rho_p-\rho_n)/\rho$. For small values of $I$, one can write
\begin{eqnarray}
E(I)/A=E(0)/A+\epsilon_1 I^2+O(I^3)
\end{eqnarray}
where $\epsilon_1$ is the asymmetry energy coefficient. This coefficient
must be positive to guarantee the stability of symmetric nuclear
matter, otherwise a small fluctuation of $I$ would amplify 
leading to neutron matter as the more stable form of nuclear
matter. 
In Figure\ref{fig3} is displayed the asymmetry energy $\epsilon_1$ as a 
   function of the density for the same interactions used in Figure 1. 
   The value of $\epsilon_1$ as deduced from the semi-empirical mass
   formula ($\epsilon \simeq 32$ MeV at saturation, indicated in
   Figure 2 as a small black rectangle) is reproduced by most of the
   effective interactions.
   Apart from the
parametrization SLy230, the displayed Skyrme interactions lead
to an isospin instability for densities a few times higher than
the saturation density. The onset of such an instability is
the value of the density at which $\epsilon_1$ becomes zero.    

\begin{figure}
\centering
\includegraphics[scale=0.7,angle=-90]{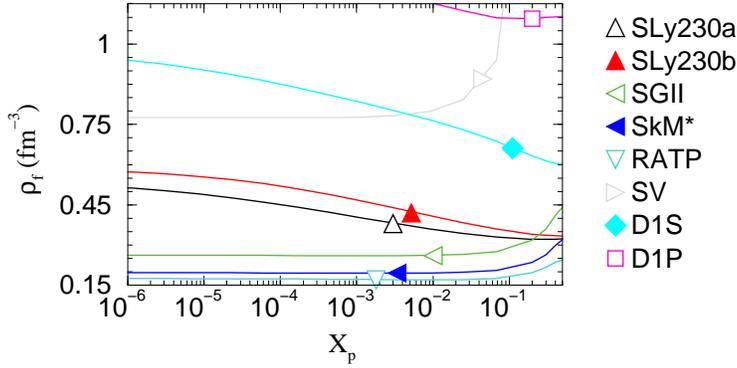}
\caption{Ferromagnetic density as a function of the proton
fraction $X_p$ for a selected set of Skyrme and Gogny effective
interactions.}
\label{fig4}
\end{figure}

We have also looked at the spin properties of the effective
interactions. In particular, we have calculated the magnetic
susceptibility of asymmetric nuclear matter~\cite{jeje}. In
Figure~\ref{fig4} is plotted as a function of the proton
fraction $X_p=\rho_p/\rho$ the value of the ferromagnetic
density, which is defined as the density at which a
ferromagnetic instability appears. It can be seen 
that all Skyrme interactions predicts such an instability in the
range of densities of relevance for the study of a proto-neutron
star. Note that the case of pure neutron matter corresponds to
$X_p=0$, at the left of the figure. In contrast, Gogny-type
interactions predict a ferromagnetic instability for densities
which are well beyond the range of values where the mean field
plus effective interactions can be reasonably used.

In conclusion, we shall employ the interactions SLy230b and D1P
for our calculations of the dynamical structure factors in neutron matter.

\section{Dynamical structure factors}

\begin{figure}
\centering
\includegraphics[scale=0.6,angle=-90]{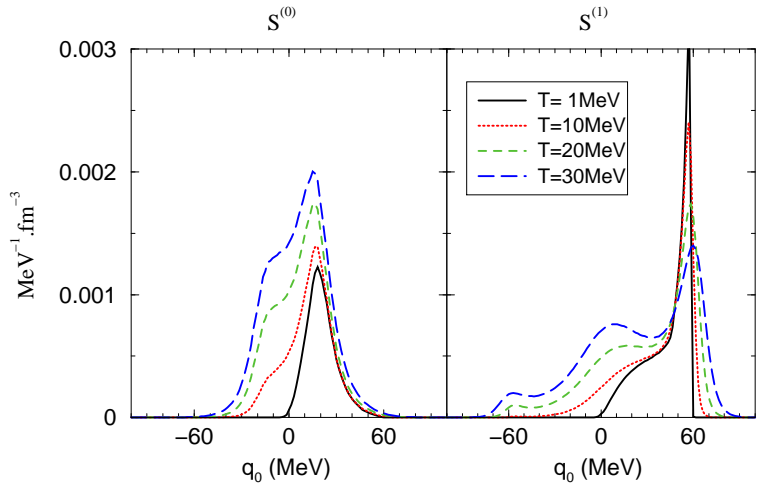}
\caption{RPA dynamical structure factors ${\cal S}^{(S)}(q,T)$
for neutron matter as a function of transferred energy $q_0$.
Calculations have been done at the saturation density of
symmetric nuclear matter, for a transferred momentum 
${\rm q}=k_F/3$ and for several temperatures. The interaction is
SLy230b.}
\label{fig5}
\end{figure}

The detailed balance relation allows one to obtain the dynamical
structure factors ${\cal S}^{(S)}$ from the imaginary part of
the response function $\chi^{(S)}(q,T)$ in pure neutron matter as 
\begin{eqnarray}
{\cal S}^{(S)}(q,T)=-\frac{1}{\pi}\frac{1}{1-\exp(-{\rm q}_0/T)} 
\,\,\,\Im {\rm m} \,\,\chi^{(S)}(q,T)~.
\end{eqnarray}
We recall that $q$ is a four-vector $({\rm q}_0, {\bf q})$, and this
expression is valid for both positive and negative transferred energies
${\rm q}_0$. 

For Skyrme type interactions, the response function can be
exactly calculated including the full RPA correlations (see
\cite{NHV:99} and references therein). 
The expressions for pure
neutron matter are given in \cite{NHV:99}
(note that, in \cite{NHV:99} Eq.(20)
should be multiplied by a factor $(1-e^{-\beta\omega})/2$). 
In Figure~\ref{fig5}
are displayed the dynamical structure factors for the density
($S=0$) and spin-density ($S=1$) channels as a function of the
transferred energy ${\rm q}_0$ from the probe to the medium.
At zero temperature, only positive values of the transferred energy are
accessible, which accounts for the heating of the medium by the
probe. As the temperature increases, the structure factors 
grow in the negative transferred energy domain. 
This behaviour corresponds to the
cooling of the medium by the probe. The pronounced zero sound in the
spin-density channel at zero temperature is damped when
the temperature increases. As a general rule, when the value of the
density increases and approaches the ferromagnetic point
mentioned in the previous section, a dramatic peak appears in
the structure factor~\cite{NHV:99,jeje}.

We have also calculated the response function assuming several
interesting approximations for the particle-hole effective
interaction. For a free Fermi gas  
the response function is given by the well-known
Lindhard function. The effects of the interaction at the
mean-field or Hartree-Fock level imply the use of an effective
mass in the Lindhard function. This is exact for Skyrme-Hartree-Fock and
represents a convenient approximation in the case of
Gogny-Hartree-Fock\cite{jeje}. 
The particle-hole interaction
may be also approached by a Landau form, containing only 
$l=0$ multipole, or $l=0, 1$ multipoles. We stress that in all cases the
particle-hole interaction has been consistently obtained from
the particle-particle interaction used to describe the EOS.
The full RPA response function has been exactly calculated for
Skyrme interactions. For the Gogny interaction we have used the effective
mass approximation for the Hartree-Fock response and the Landau
approximation for the RPA response. 

\section{Neutrino mean free paths}

\begin{figure}
\centering
\includegraphics[scale=0.7,angle=-90]{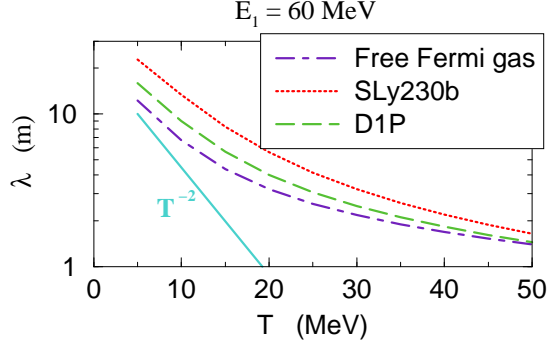}
\caption{Mean free path in neutron matter as a
function of the temperature. The curves labeled SLy230b and D1P
have been obtained within the mean field approximation for the
response function. The energy of the incident neutrino
is $E_1=60$ MeV.} 
\label{fig6}
\end{figure}

The temperature dependence of the neutrino mean free path can be
seen in Figure~\ref{fig6}, where the calculated $\lambda$ has
been represented in the case of a free Fermi gas, and in the mean
field approximation using SLy230b and D1P interactions. We have
chosen an incoming neutrino energy $E_1=60$ MeV. Only at very low
temperatures the behavior of $\lambda$ follows the expected law in
$T^{-2}$~\cite{RPL:98} (indicated also in the figure). As
the temperature increases the interaction effects, contained in the
effective mass, become less important and one reaches the free Fermi gas value
at very high temperatures. 

\begin{figure}
\centering
\includegraphics[scale=0.8,angle=-90]{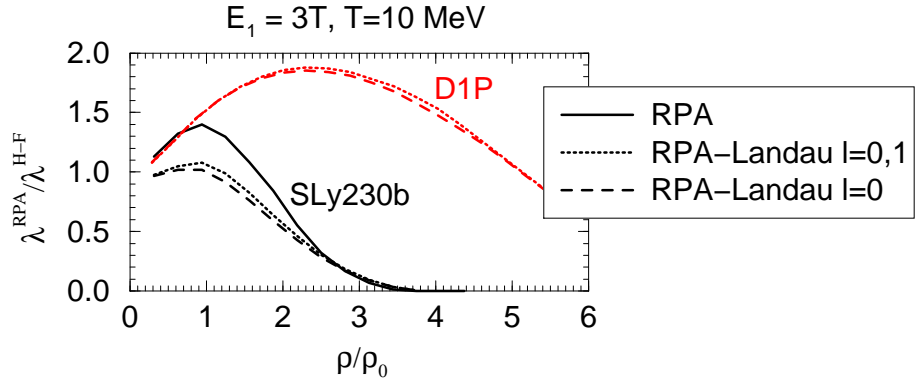}
\caption{The relative mean free path $\lambda^{\rm RPA}/\lambda^{\rm HF}$
as a function of the density for interactions SLy230b and D1P.}
\label{fig7}
\end{figure}

The effects of the residual interaction, i.e., of RPA correlations, can be
seen in 
Figure~\ref{fig7}, where we have represented the 
ratio of RPA to Hartree-Fock mean free paths 
as a function of the density. For SLy230b, we
have calculated the response function using either the full
particle-hole interaction (labeled RPA in the figure) or its
Landau approximation (labeled RPA-Landau in the figure). Only
the latter approximation has been used for the interaction D1P. 

In the case of interaction SLy230b one can see that the effects
of the residual interaction in Landau approximation (either
$l=0$ or $l=0,1$) do not produce any significant difference as
compared to the HF approximation for densities around the
saturation value. In the same region, the full residual
interaction produces a sizeable increase of $\lambda$. All three solutions
predict a strong decrease of the mean free path when the density
reaches the ferromagnetic density. This is due to the very strong
axial response function as the system comes closer to the
ferromagnetic instability. In contrast, for interaction D1P a
significant difference with respect to the HF results appears 
when the residual interaction is considered at the Landau level.
Note that the differences in the results obtained using the 
Landau approximations with $l=0$ or with $l=0,1$ are quite small. 
Finally, the force D1P also
predicts a vanishing mean free path when the density increases, but
the ferromagnetic instability happens at very high densities,
well beyond the domain of applicability of the effective interaction
approach. 

\section{Conclusions}
In the domain of densities and protonic fractions that are relevant to the
proto-neutron star evolution, the use of effective nucleon-nucleon
interactions which are originally designed for describing the neighbourhood
of normal nuclear matter is a delicate issue. These interactions may fail to
reproduce the EOS of neutron matter (SIII. SV, D1, D1S), or they lead to
isospin and/or spin instabilities at 2-3 times normal density (Skyrme forces).

Having this in mind, it is still possible to use some of the existing
effective interactions to calculate consistently the nuclear response
functions and hence the neutrino mean free paths in neutron matter. The
evolution of the response functions with increasing temperatures shows that
neutrinos can efficiently mediate the cooling process of hot neutron matter.
We find that RPA correlations have a significant effect on mean free paths.
With Skyrme interactions, RPA correlations tend to increase mean free paths
at densities around normal density whereas at larger densities they predict
very short mean free paths because of the onset of spin instabilities. Thus,
these interactions do not favor a fast cooling process. On the other hand,
the Gogny interaction predicts a quite different scenario where RPA
correlations increase mean free paths in the whole range of densities (up to
about $5\rho_0$).  
In this latter case where 
ferromagnetic instabilities are not present, the mean field and the RPA
correlations make the nuclear matter more transparent compared
to the free Fermi gas model. Thus, neutrinos would escape more quickly 
and the delayed mecanism will be less efficient.

%

\end{document}